\documentstyle[12pt]{article}
\title{Impurity Effects on Quantum Depinning of\\ 
Commensurate Charge Density Waves}

\author{ Masanori {\sc Yumoto}\footnote{E-mail: yumoto@presto.phys.sci.osaka-u.ac.jp},
Hidetoshi {\sc Fukuyama}$^{1}$, Hiroshi {\sc Matsukawa}  
\\ and Naoto {\sc Nagaosa}$^{2}$ }
\date{(recieved April 14, 2000)}
\begin{document}
\maketitle
\begin{center}
Department of Physics, Osaka University, Toyonaka, Osaka 560-0043\\
$^1$ Institute for Solid State Physics, University of Tokyo, Roppongi, 
Tokyo 106-8666\\
@$^2$ Department of Applied Physics, University of Tokyo, Hongo, Tokyo 113-8656 
\end{center}

\begin{abstract}
We investigate quantum depinning of the 
one-dimensional (1D) commensurate charge-density wave (CDW) in the presence of 
one impurity theoretically.  
Quantum tunneling rate below but close to 
the threshold field is calculated at absolute zero temperature 
by use of the phase Hamiltonian within the WKB approximation. 
We show that the impurity can induce localized fluctuation 
and enhance the quantum depinning. 
The electric field dependence of the tunneling rate in 
the presence of the impurity is different from that in its absence. 
\end{abstract}


\sloppy

\section{Introduction}

The tunneling of macroscopic object, the so called Macroscopic Quantum 
Tunneling (MQT), is one of the most dramatic quantum effects.
There has been many theoretical studies of MQT.~\cite{rf:Weiss}
Most of them are dealing with regular systems.
In many actual cases, however, some kind of randomness exists and is 
expected to play an important role.
Here we study the effects of randomness for the case of charge-density wave 
(CDW) as a typical example.~\cite{rf:G}
MQT of the commensurate CDW below but close to the threshold field 
was investigated by Nakaya and Hida (NH).~\cite{rf:Nakaya}
They employed the WKB approximation in order to obtain the tunneling rate 
as a function of the electric field.
However the effects of randomness were not considered there.
In the framework of classical mechanics it is shown that the randomness can 
make lower the threshold field for the depinning of the commensurate CDW with 
one impurity.~\cite{rf:Yumoto}
In this paper we calculate the quantum tunneling rate of 
the commensurate CDW with one
impurity. The 1D phase Hamiltonian~\cite{rf:HF1}~\cite{rf:HF2}~\cite{rf:LR} is adopted,
and the electric
field dependence of the tunneling rate is
obtained by use of Langer's method~\cite{rf:Langer} at absolute zero.
It is found that the impurity can induce localized fluctuation and 
enhance the tunneling rate near the threshold field. 
We have different electric field dependence of 
the tunneling rate in the presence of the impurity from that given 
by NH in its absence.~\cite{rf:LT22}

\section{The Model and the Threshold Field}

The CDW is the modulation of the charge density, 
$\rho(X, T)=\rho_0\cos\left(2k_{\rm F}X+\phi(X, T)\right)$, 
where $\rho_0$ is the amplitude of the modulation, $\phi$ is its phase  
and $k_{\rm F}$ is 
the Fermi wavenumber. 
We investigate the 1D commensurate CDW 
with one impurity located at $X_i$ at absolute zero temperature.   
The action of the system is described by $\phi$ as 
\begin{eqnarray*}
S = \int {\rm d}T\int {\rm d}X
\left[ A\left(\frac{\partial\phi}{\partial T} \right)^2  
-B\left(\frac{\partial\phi}{\partial X} \right)^2 
+ F_{\rm ele}\phi \right. \hspace{5cm} \mbox{}
\end{eqnarray*} 
\begin{equation}
\hspace{3cm} \mbox{}
 - g \left(1 - \cos(M \phi) \right) 
+  V\cos(2 k_{{\rm F}} X + \phi)
\delta(X-X_{i})  \mbox{\Huge ]}.
\label{eq:Eaction}
\end{equation} 
The first term in the integrand is the kinetic energy density with   
$A=(\hbar v_{{\rm F}})/(4 \pi v_{\phi}^2)$, where $v_{{\rm F}}$ is the Fermi velocity and 
$v_{\phi}$ is the phason velocity. The second is the elastic one with 
$B=(\hbar v_{{\rm F}})/(4 \pi)$ and the third is the electric field one. Here 
$F_{\rm ele} = ({\rm e}E_{\rm ext})/(2 \pi)$ and ${\rm e}<0$ is the charge of an electron and 
$E_{\rm ext}$ is the electric field. The fourth represents the energy resulting from the commensurability, $M \equiv \pi/(k_{\rm F} a)$,  and $a$ is a lattice constant. 
The last is the impurity potential energy with the amplitude $V$. 
In this action there exist the characteristic length and time,   
$\xi = \sqrt{(2 B)/(M^2 g) }$ 
and $\xi_{t} = \sqrt{(2 A)/(M^2 g) }$, which are
the phase coherence length and time 
due to the commensurability, respectively. 
We scale this action by $\xi$ and $\xi_{t}$, i.e.
$x =X/\xi$, $t = T/\xi_{t}$. The Euclidean action is introduced as 
$S_{\rm E}=iS$;
\begin{eqnarray*}
S_{\rm E} =C \int {\rm d}\tau \int {\rm d}x \left[  
\frac{1}{2}\left(\frac{\partial\phi}{\partial \tau} \right)^2 
+\frac{1}{2}\left(\frac{\partial\phi}{\partial x} \right)^2 
- \varepsilon\phi \right. \hspace{5cm} \mbox{} \nonumber
\end{eqnarray*}
\begin{equation}
\mbox{} \hspace{3cm}
\left. + \frac{1}{M^2}\left(1 - \cos(M \phi) \right)
- v\cos(\chi(x_{i}) + \phi)\delta(x-x_{i}) \right],  
\label{eq:Lag}
\end{equation}
where $\tau=-it$ is the imaginary time, $C=M^{2}g\xi\xi_t=2\sqrt{A B}$, 
$\varepsilon = F/(g M^2)$, and $v = V/(g \xi M^2)$. 
The typical magnitude of $C$ is given as
\begin{equation}
C/\hbar=\sqrt{m^{\ast}/m_{\rm b}}/ 2 \pi \simeq 2 \sim 2\times 10. 
\label{eq:ch}
\end{equation} 
Here 
$m^{\ast}$ and $m_{\rm b}$ are the effective mass of the CDW and the band mass, 
respectively.  
We define $x_{i}=X_{i}/\xi$ and 
$\chi(x_{i})= (2 \pi z)/M $, where  
$z=X_{i}/a $ characterizes the location of 
the impurity relative to the crest of the undeformed CDW with minimum  
commensurability energy.   
The range of $\chi$ can be set as $-\pi/M \leq \chi \leq \pi/M$, 
because the action except of the impurity term 
has the periodicity of $2\pi/M$ 
with respect to the phase variable. 
We consider the case of $v \geq 0$ hereafter.

%\section{Classical Threshold Field}

We first consider the threshold field at absolute zero temperature 
in the framework of classical static mechanics.~\cite{rf:Yumoto} 
The first variational equation of the action,  
\begin{equation}
- \phi'' - \varepsilon + \frac{1}{M}\sin(M\phi) 
+ v\sin(\chi + \phi)\delta(x-x_{i}) =0,  
\label{class}
\end{equation}
determines 
the ground state configuration of the phase with   
the boundary conditions,  
$\phi(\pm \infty)=\phi_0$ and  
$\phi'(\pm \infty)=0$. Here
\[
\phi_0 \equiv \frac{1}{M}\arcsin(\varepsilon M)  
\nonumber
\]
is the ground state configuration in the absence of the impurity.  

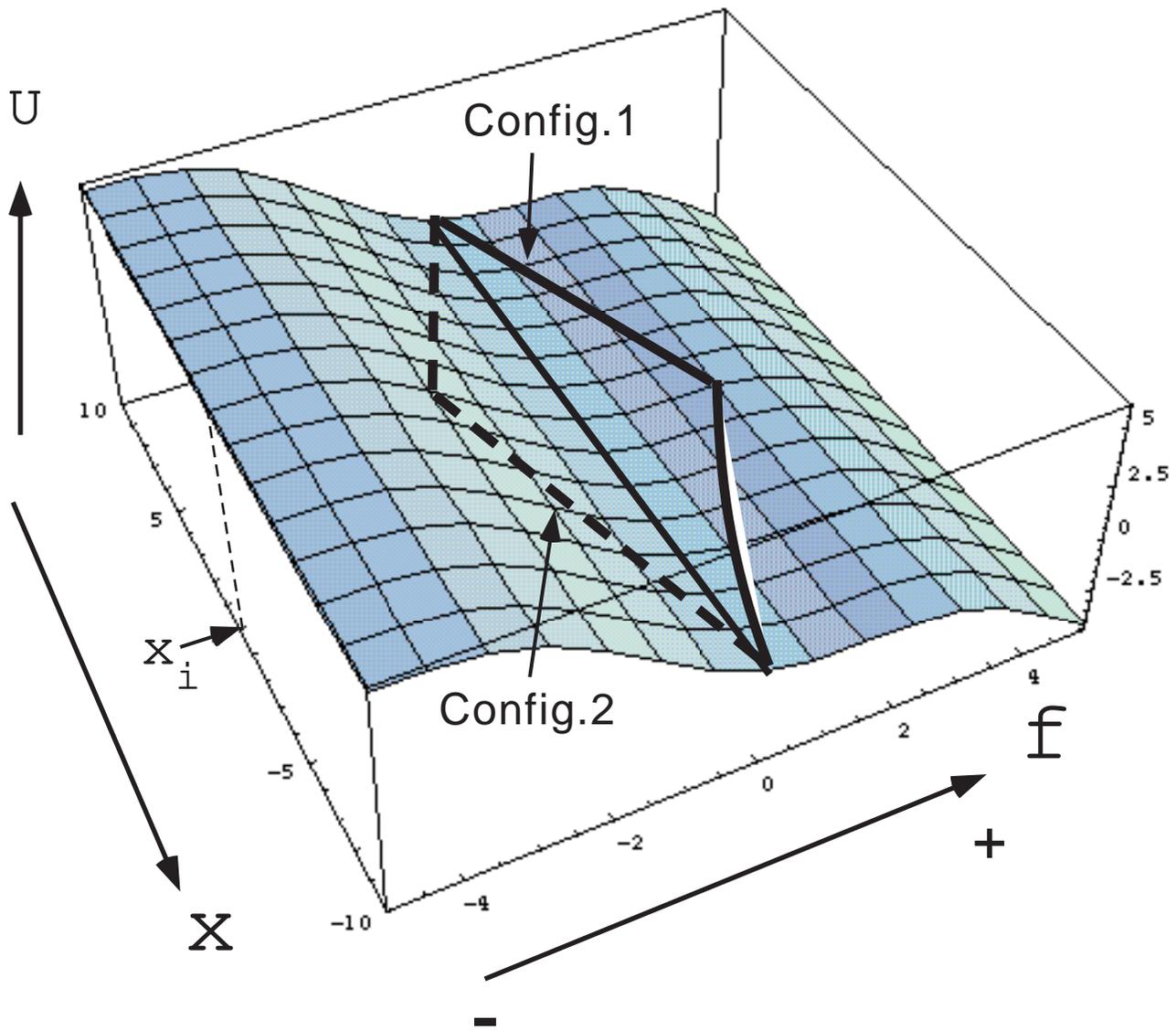
\begin{figure}[ht]
%\vspace{10cm}
%\epsfigure{file=figure1.eps,height=10cm}
\caption{A schematic picture of two kinds of the configurations, Config.1 and 2,  
with respect to $\chi$.  The potential in the absence of the impurity is 
$U=-\varepsilon \phi + (1-\cos\phi)/M^2$. }
\label{fig:conf}
\end{figure}

Depending on the range of $\chi$, there are two kinds of configurations. 
If $\chi$ is in the range of 
$-\pi/M \leq \chi \leq 0$, the solution is located in the positive side (Config.1)  
as shown in Fig.~\ref{fig:conf}.  
If $\chi$ is in the range of $0 \leq \chi \leq \pi/M$, the solution 
is located in the negative side (Config.2). 
We take electric field $\varepsilon \geq 0$. Under this condition, Config.1   
has more tendency to depin than Config.2. 
Therefore, we consider only the case of 
$-\pi/M \leq \chi \leq 0$, because we are interested in the lowering of 
the threshold field. 

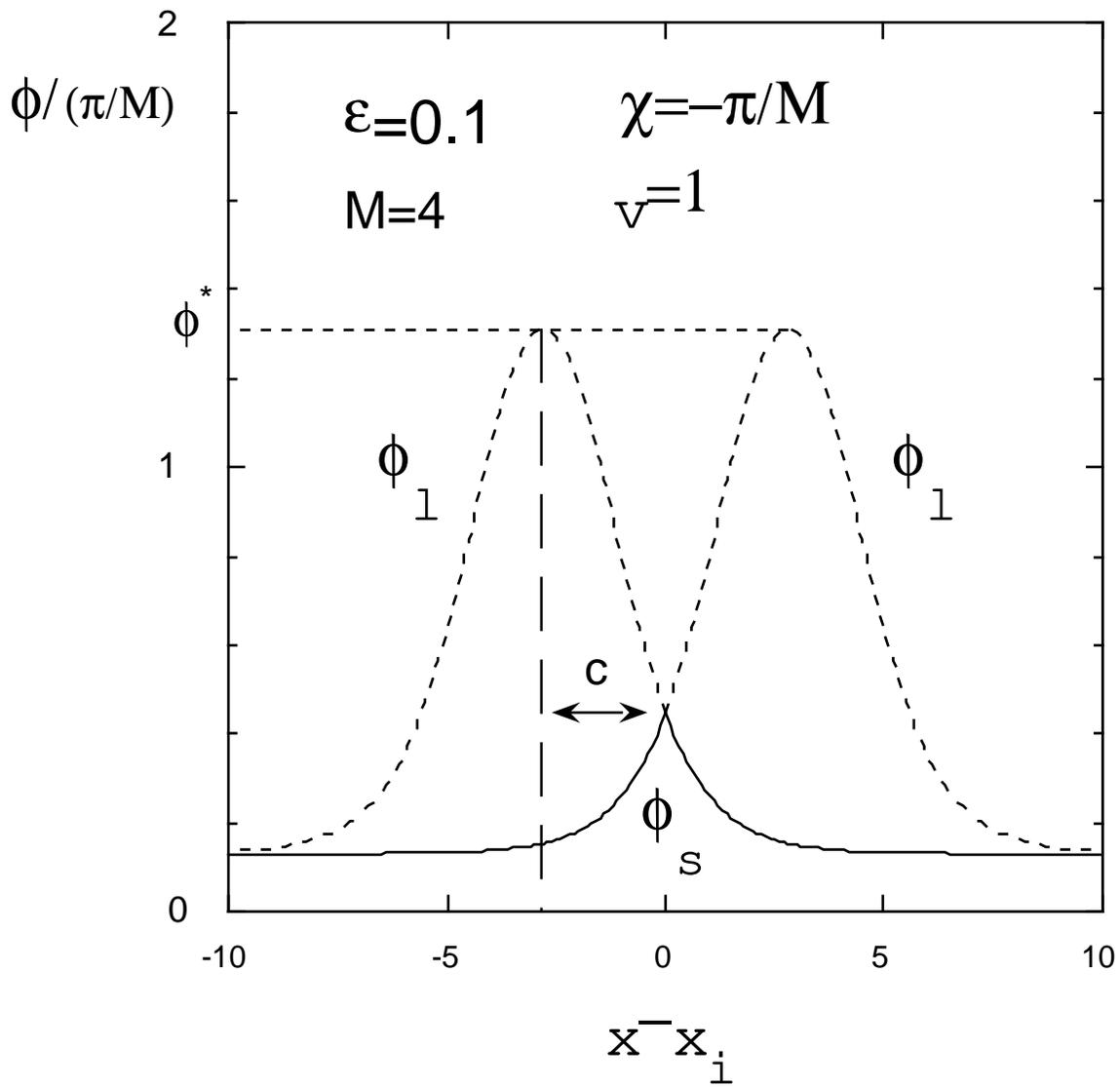
\begin{figure}[ht]
%\vspace{10cm} 
%\epsfigure{file=figure2.eps,height=10cm}
\caption{The ground state configuration of the phase, $\phi_s(x)$, in the presence 
of the impurity. The solid line is $\phi_s(x)$ for a choice of  
 $M=4$, $v=1$, $\chi=-\pi/M$ and $\varepsilon = 0.1$. }
\label{fig:2}
\end{figure}

The ground state solution, $\phi_{s}(x)$, 
in the presence of the impurity 
which is determined by eq.~(\ref{class}) is obtained by the connection 
at the impurity site  
of two non-uniform solutions in the absence of the impurity, $\phi_{l}(x)$
(Fig.~\ref{fig:2}).  
The non-uniform solution, $\phi_{l}(x)$, is given by 
\begin{equation}
\phi_{l}' = -\left[ 2 \left[ \varepsilon \left( \frac{1}{M} 
\arcsin(\varepsilon M) - 
\phi_{l} \right)
+ \frac{1}{M^2}\left(\sqrt{1-(\varepsilon M)^2} 
- \cos(M \phi_{l})\right)\right]\right]^{\frac{1}{2}},  
\label{eq:3.5}
\end{equation}
which is derived by integration of  
\begin{equation}
- \phi'' - \varepsilon + \frac{1}{M}\sin(M \phi)=0, 
\label{eq:3.1} 
\end{equation}
with the boundary condition, 
$\phi(\pm \infty)=\phi_0$ and 
$\phi'(\pm \infty)=0$. 
Integration of eq.~(\ref{eq:3.5}) is 
obtained only by numerical treatment. When we carry out 
the integration practically, we need the information of the phase at 
the impurity site obtained analytically as follows.
The ground state solution, $\phi_{s}(x)$, is given by 
\begin{equation}
\phi_{s}(x) = \phi_{l}(|x-x_{i}| + c) ,  
\label{classsol}
\end{equation}
where $c$ represents the distance between the center of $\phi_{l}$ and 
$x_i$.  
Substituting eq.~(\ref{classsol}) into eq.~(\ref{class}), we obtain  
\begin{equation}
- 2 \phi_{l}'(c) + v \sin(\chi + \phi_{l}(c)) = 0. \label{cla1}
\end{equation}
By use of eq.~(\ref{eq:3.5}), eq.~(\ref{cla1}) is rewritten as 
\begin{eqnarray*}
\left[ 2 \left[ \varepsilon \left( \frac{1}{M} \arcsin(\varepsilon M) 
- \phi_{l}(c) \right)
+ \frac{1}{M^2}\left(\sqrt{1-(\varepsilon M)^2} 
- \cos(M \phi_{l}(c))\right)\right]\right]^{\frac{1}{2}} \nonumber 
\end{eqnarray*}
\begin{equation}
\mbox{} \hspace{3cm} + \frac{1}{2}v\sin(\chi + \phi_{l}(c)) = 0.    
\label{result}
\end{equation}
This equation determines the phase, $\phi_{l}(c)=\phi_s(x_i)$,  
at the impurity site of 
the ground state solution, instead of $c$, when 
$\varepsilon$ is given. 
Substituting this $\phi_s(x_i)$ into eq.~(\ref{eq:3.5}), we have the derivative at 
$x_i$, and then accomplish the numerical integration with the boundary conditions 
at $x_i$. 

%\subsection{Classical Threshold Fields}

The threshold field is   
determined as the field where the ground state solution 
becomes unstable. 
The eigenvalue equation of the fluctuations, $\delta \phi(x)$, 
around the ground state solution, $\phi_{s}(x)$, is 
expressed as 
\begin{equation}
\left[ - \frac{\partial^2}{\partial x^2} + V_{s}(x) \right]\delta \phi 
= \lambda \delta \phi,  
\label{fluc}
\end{equation}
where 
\begin{equation}
V_{s}(x) \equiv  
\cos\left(M \phi_{s}(x)\right)
+ v\cos\left(\chi + \phi_{s}(x)\right)\delta(x-x_{i}),  
\label{eq:schpot}
\end{equation}
and $\lambda$ is the eigenvalue. 
The threshold field is determined by the condition that  
the lowest $\lambda=0$.  
In the absence of the impurity, 
$V_s(x)$ becomes 
\begin{equation}
\Delta_{\rm g}=\sqrt{1-\left(\varepsilon M\right)^2}. 
\end{equation}
Then the threshold is $\varepsilon_{\rm T}=1/M$. 
At $\varepsilon_{\rm T}$  
the barrier of the potential energy density in the absence of the impurity, 
\begin{equation}
U(\phi)=-\varepsilon\phi+\frac{1}{M^2}(1-\cos(M\phi)), 
\label{eq:U}
\end{equation}  
disappears. 
The threshold field, $\varepsilon_{\rm c}$, 
in the presence of the impurity~\cite{ref:note1}   
is lower than that in its absence 
under the condition of $- \pi/M \leq \chi < - \pi/(2M)$ 
and $v>0$. 
When $v$ is fixed, $\varepsilon_{\rm c}/\varepsilon_{\rm T}$ goes to 1 
as $\chi$ tends to $-\pi/(2M)$. 
We also note that the threshold field goes to a finite value,
$\varepsilon_{\infty}$,  
even if $v \rightarrow \infty$. 
For $M=4$ and $\chi=-\pi/M$, 
$\varepsilon_{\infty}/\varepsilon_{\rm T} \simeq 0.725$, which is 
the lowest threshold of the present model.

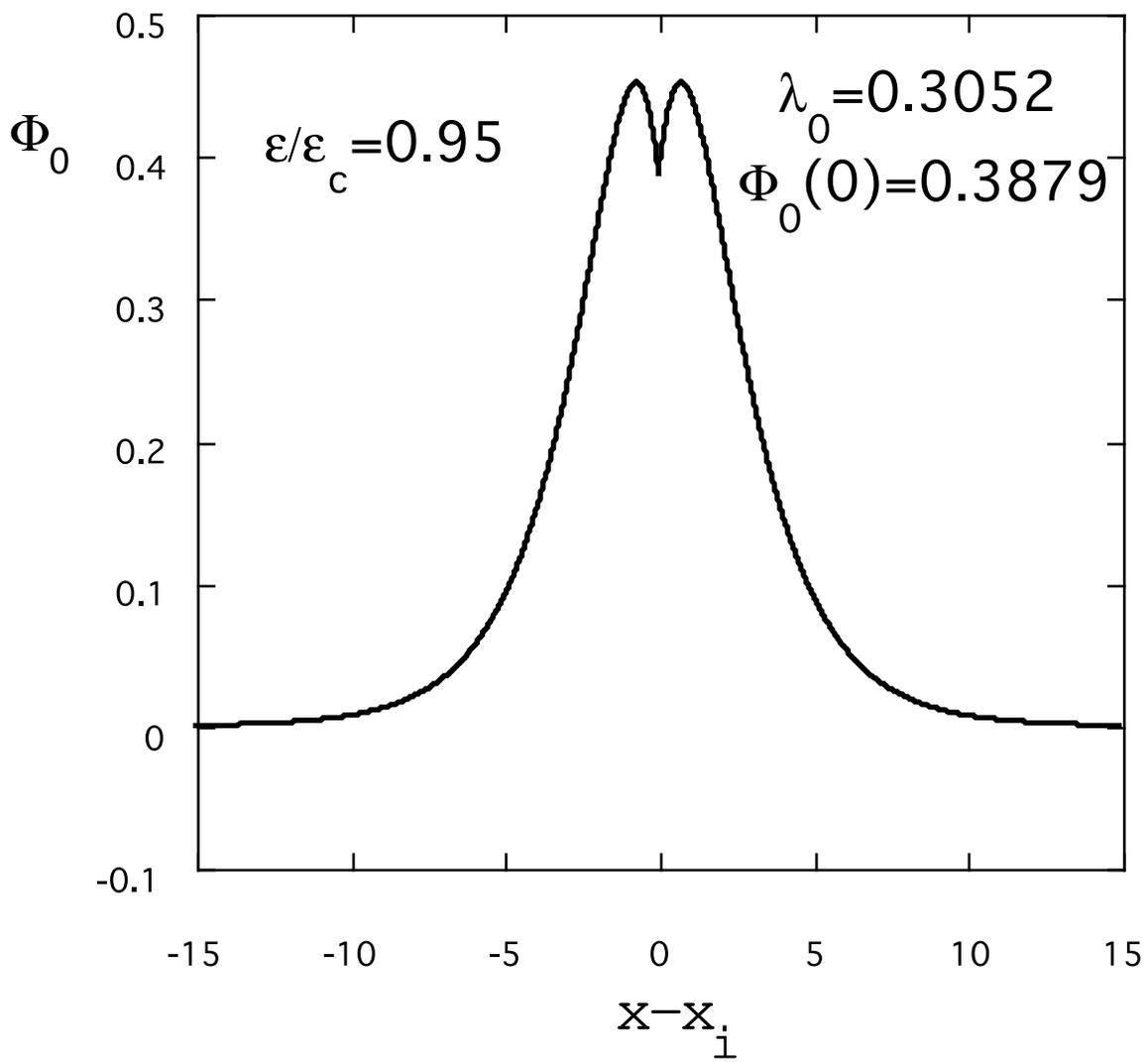
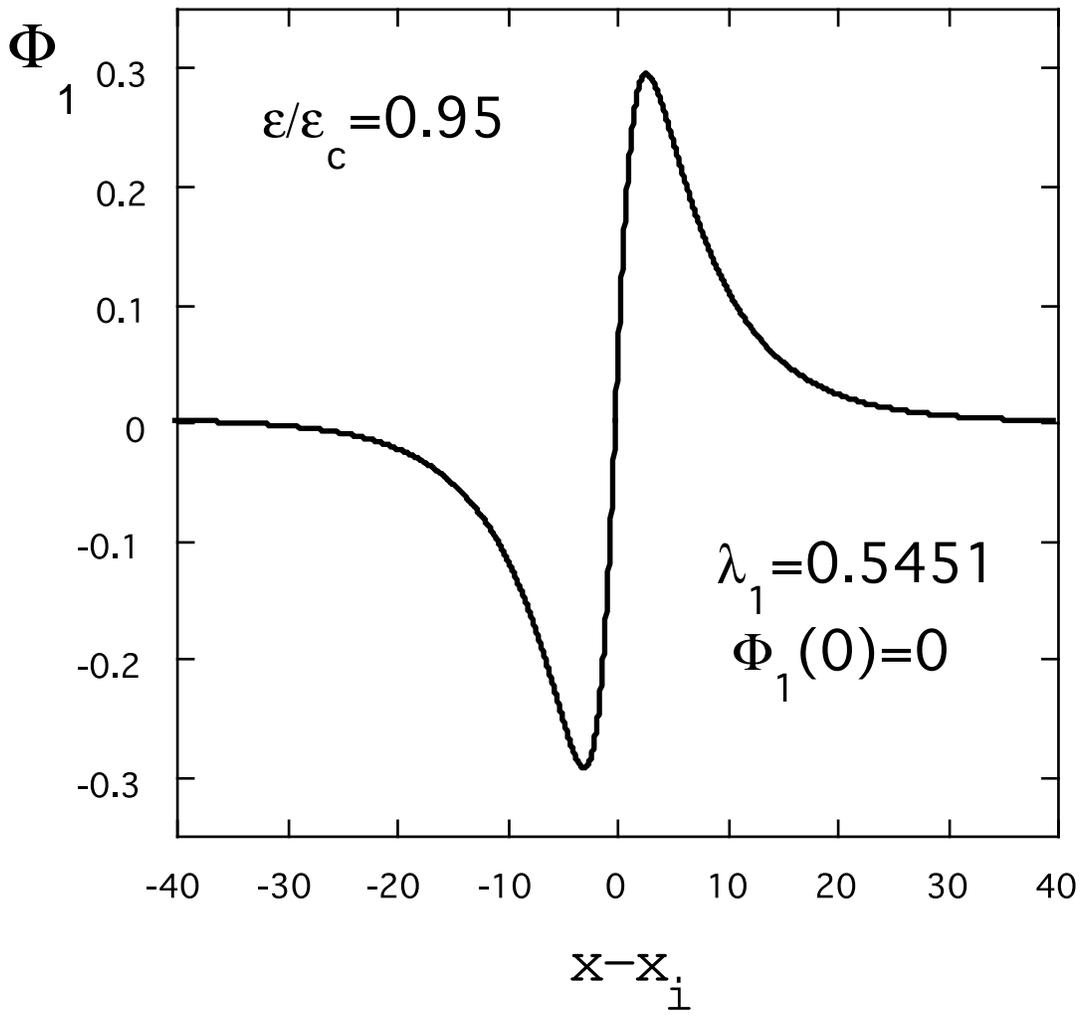
\begin{figure}[ht]
%\vspace{20cm}
%\epsfigure{file=figure4a.eps,height=10cm}
%\epsfigure{file=figure4b.eps,height=10cm}
%\epsfigure{file=figure4c.eps,height=10cm}
\caption{(a): A typical example of $V_{s}(x)$ 
for a choice of $M=4$, $\chi=-\pi/M$, $v=1$ 
and $\varepsilon/\varepsilon_{\rm c}=0.95$. (b) and (c): Normalized eigenfunctions 
of the discrete eigenvalue. 
There are only two bound solutions, $\Phi_{0}(x)$ and $\Phi_{1}(x)$.}
\label{fig:Vs}
\end{figure}

First we consider the reason of the lowering of the threshold field 
in an aspect of the fluctuation. 
Due to the existence of the impurity, the potential term of the 
eigenvalue equation, $V_{s}(x)$, is non-uniform in space 
over the length scale of $\xi$ (Fig.~\ref{fig:Vs}(a)).  
We note that in the present dimensionless notation, $\xi$ is unity. 
We solve eq.~(\ref{fluc}) numerically by 
transforming a differential equation into a difference equation (the Numeov method) 
and the shooting method on $\lambda_{n}$.~\cite{rf:comput}~\cite{rf:fort} 
As the result  
two bound solutions, $\Phi_{0}(x)$ and $\Phi_{1}(x)$,  
(Fig.~\ref{fig:Vs}(b) and (c)), are obtained at most 
under the condition of $- \pi/M \leq \chi < - \pi/(2M)$ 
and $v>0$ below but near the threshold field.  
Other solutions are non-bounded solutions whose eigenvalues form 
a continuous spectrum above $\Delta_{\rm g}$.
The eigenvalue of the bound state is smaller than 
the lowest eigenvalue, $\Delta_{\rm g}$, of the continuous spectrum.  
Then applying the electric field, the lowest 
eigenvalue, $\lambda_0$, 
of the bound state becomes zero before $\Delta_{\rm g}$ becomes zero. 
In this case the threshold field is lowered by the impurity than that 
in its absence.  Here we note that in the absence of the impurity 
$V_{s}(x)$ is uniform in space and there exists the continuous spectrum only. 
In the region of $-\pi/(2M) \leq \chi$, however, 
there is no bound state solution 
of eq.~(\ref{fluc}), and the threshold is the same as that in the absence of 
the impurity.  

The bound state corresponds to the localized fluctuation over
the length scale of 
$\xi$ around the 
impurity (Fig.~\ref{fig:Vs}(b)). 
Owing to the localized fluctuation of the lowest eigenvalue, 
the CDW starts {\it local sliding} at the threshold field. 
The onset of a sliding in the absence of the impurity accompanies 
the spatially uniform and temporally continuous 
motion of the CDW. In the presence 
of the impurity, however, the sliding takes place around 
the impurity, i.e. the local sliding. 
Once the local sliding sets in, the CDW gains 
the electric field energy, which is transferred to the kinetic 
energy. Therefore, the sliding expands to the whole system  
and the whole CDW  
completely depins eventually, when dissipation is small enough. 
In the present model dissipation is not taken into account. 
The issue of dissipation is discussed in the final section.  

%\subsection{The One-Particle Picture of the Model}

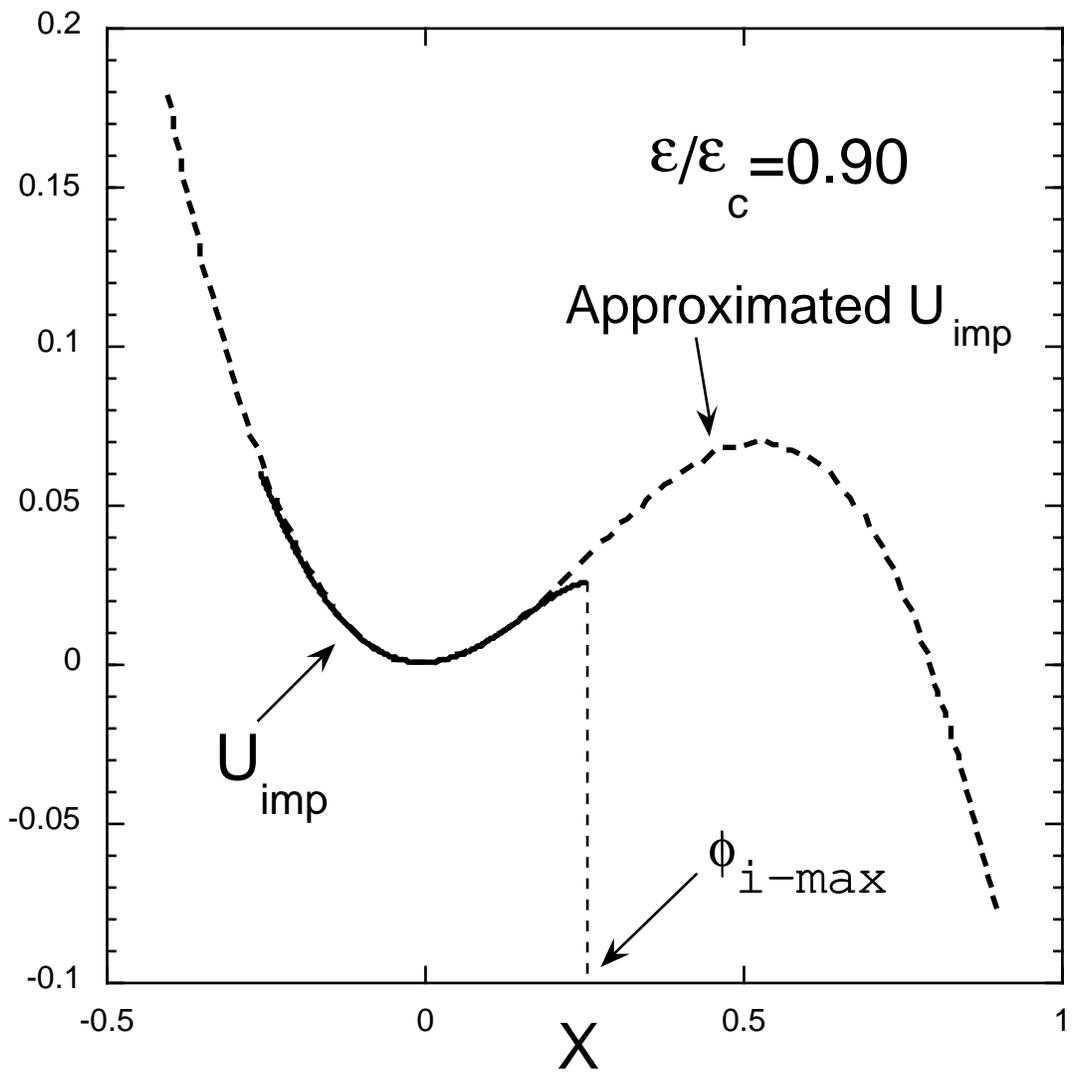
\begin{figure}[ht]
%\vspace{20cm}
%\epsfigure{file=figure3.eps,height=10cm}
\caption{ The potential curves in the presence of an impurity 
for a choice of $M=4$, $\chi=-\pi/M$, $v=1$ and 
$\varepsilon/\varepsilon_{\rm c}=0.90$. The origin of the 
potential is set at $\phi_i=\phi_s(x_i)=0$ and $U_{\rm imp}(\phi_s(x_i))=0$.  
The approximated $U_{\rm imp}$  
in terms of $X^2$ and $X^3$ ($X=\phi_i - \phi_s(x_i)$) is represented 
by the dotted line. 
For $\phi_{i}>\phi_{i-{\rm max}}$, there is no optimal configuration.}
\label{fig:curve}
\end{figure}

Next we clarify the meaning of $\varepsilon_{\rm c}$ in an aspect of potential curves  
in the 
presence of the impurity.~\cite{rf:Yumoto} 
In the absence of the impurity the phase is uniform in 
the ground state, and hence 
the potential energy curve can be expressed 
with respect to the uniform phase variable as eq.~(\ref{eq:U}). 
With the impurity, however, the 
phase is not uniform. Apparently we should consider 
the potential curve in a functional 
space in this case.  
We can reduce the problem, however, to a one-particle problem by adopting 
the phase variable at the impurity site, $\phi_{i}$, 
as a kind of generalized collective coordinate.  
For each value of $\phi_{i}$, 
we can determine the 
lowest energy configuration; we solve eq.~(\ref{eq:3.1}) 
under the boundary condition, 
$\phi(x_{i})=\phi_{i}$ and 
\[
\phi'(x_{i}) = -\left[ 2 \left[ \varepsilon 
\left( \frac{1}{M} \arcsin(\varepsilon M) - 
\phi_{i} \right)
+ \frac{1}{M^2}\left(\sqrt{1-(\varepsilon M)^2} 
- \cos(M \phi_{i})\right)\right]\right]^{\frac{1}{2}},  
\]
which is just the same as eq.~(\ref{eq:3.5}) at the impurity site. 
Substituting this optimal configuration into the Euclidean Lagrangian,  
we obtain the potential energy, $U_{\rm imp}(\phi_i)$, as 
the function of $\phi_{i}$; 
\begin{equation}
U_{\rm imp}(\phi_i)=-2\sqrt{2}\int^{\infty}_{\phi_i} {\rm d}\phi 
\sqrt{E_0(\phi)} - v\cos(\chi+\phi_i), 
\label{eq:uimp}
\end{equation}
where 
\[
E_0(\phi)=\varepsilon 
\left( \frac{1}{M} \arcsin(\varepsilon M) - 
\phi \right)
+ \frac{1}{M^2}\left(\sqrt{1-(\varepsilon M)^2} 
- \cos(M \phi)\right). 
\] 
The value of $\phi_i$ that minimizes 
$U_{\rm imp}(\phi_i)$ is $\phi_s(x_i)$. 
Then we introduce a variable, $X \equiv \phi_i - \phi_s(x_i)$, and consider 
$U_{\rm imp}$ as a function of $X$ hereafter.   
The curves of $U_{\rm imp}(X)$ is shown 
in Fig.~\ref{fig:curve}. For each electric field there is $\phi_{i-{\rm max}}$, and 
for $\phi_{i}>\phi_{i-{\rm max}}$ no such an optimal configuration exists.
Therefore once the barrier disappears at $\varepsilon_{\rm c}$, 
the sliding sets in  
because of the absence of any stable configuration, and then the CDW starts to depin.   
The value of $\phi_{i-{\rm max}}$ is same as the maximum value 
of $\phi_l(x)$, $\phi^{\ast}$ (see Fig.~\ref{fig:2}).

\section{Quantum Depinning in the Presence of an Impurity}

We derive the effective action based on the one-particle picture 
near the threshold field, $\varepsilon_{\rm c}$,  
and calculate the tunneling rate in the presence of the impurity 
within the WKB approximation 
by Langer's method.~\cite{rf:Langer} 
The tunneling rate is enlarged and its electric field 
dependence is changed by the impurity 
from that in its absence.~\cite{rf:LT22} 

\subsection{The Effective Action near the Threshold Field}

To investigate quantum tunneling problems theoretically by the semiclassical 
approximation, we must choose a suitable 
tunneling process, or in other words, the bounce solution. If we can derive 
the effective potential as a function of a single variable, the bounce solution is 
easily obtained by solving the equation of motion of a single degree of freedom. 
We call such a variable as the {\it tunneling variable}. 
In the present case the tunneling variable is 
the phase variable at the impurity site, $\phi_{i}$ or $X$.   
The potential curve has been already obtained as a function 
of the tunneling variable in 
eq.~(\ref{eq:uimp}). 

In order to derive the kinetic term, 
we follow the method  used to 
derive the effective action in the Tomonaga-Luttinger model 
with a single barrier.~\cite{rf:Furusaki} 
We introduce the partition function of the present model as 
\begin{equation}
Z=\int {\cal D}\phi \exp\left(-S_{\rm E}\left[\phi \right]/\hbar\right)
\label{eq:partition}
\end{equation}  
where $S_{\rm E}[\phi]$ is 
the Euclidean action which is given by eq.~(\ref{eq:Lag}). 
The phase, $\phi(\tau, x)$, is expressed by the sum of the static part, $\phi_{s}(x)$, which 
is the solution of eq.~(\ref{class}), and the  
fluctuation part, $\delta\phi(\tau, x)$; 
\begin{equation}
\phi(\tau, x) = \phi_{s}(x) + \delta\phi(\tau, x). 
\label{eq:psf}
\end{equation}
Substituting eq.~(\ref{eq:psf}) into the action and expanding it in  
powers of $\delta\phi(\tau, x)$,  
we obtain the action up to the second order of $\delta\phi(\tau, x)$ as
$S_{\rm E} \simeq S_{0}+S_{2}$, 
where 
\[
S_{0}= C\int {\rm d}\tau{\rm d}x \left[
\frac{1}{2}\left(\frac{\partial\phi_{s}(x)}{\partial x} \right)^2 
+ V(\phi_{s}(x))\right], 
\]
\begin{equation}
S_{2}= C\int {\rm d}\tau{\rm d}x \;
\left[\delta\phi(\tau, x)\frac{1}{2}\left[
-\frac{\partial^2}{\partial \tau^2} 
-\frac{\partial^2}{\partial x^2} 
+ V''(\phi_{s}(x))\right]\delta\phi(\tau, x)\right]. 
\label{eq:S2}
\end{equation}
Here 
\begin{equation}
V(\phi) = - \varepsilon\phi 
+ \frac{1}{M^2}\left(1 - \cos(M \phi) \right)
- v\cos(\chi + \phi)\delta(x-x_{i}). 
\label{eq:POT}
\end{equation}
In order to integrate the phase, $\delta\phi(\tau, x)$, under the condition, 
$\delta\phi(\tau, x_{i})=\phi_i( \tau)-\phi_s(x_i) = X(\tau)$,   
auxiliary fields, $\Lambda(\tau)$, are introduced and the identity,  
\[
1=\int {\cal D}X(\tau) \int {\cal D}\Lambda(\tau) 
\exp\left[ i \int {\rm d}\tau \Lambda(\tau) 
\left(\delta\phi(\tau, x_{i})-X(\tau)\right)\right], 
\label{eq:hojo}
\]
is employed. 
Then the partition function is expressed as 
\begin{eqnarray*}
Z&\simeq&\exp\left(-S_{0}/\hbar\right) 
\int {\cal D}\delta\phi(\tau, x) 
\int {\cal D}X(\tau) \int {\cal D}\Lambda(\tau)  \\
&\times&\exp\left[-S_{2}/\hbar + i C/\hbar \int {\rm d}\tau \Lambda(\tau) 
\left(\delta\phi(\tau, x_{i})-X(\tau)\right)\right].
\end{eqnarray*}
We diagonalize $S_{2}$ by use of the eigenfunction; 
\[
\delta\phi(\tau, x)=\sum_{m,n} c_{m,n} \exp\left(i\omega_{m}\tau\right) 
\Phi_{n}(x), 
\]
where $\Phi_{n}(x)$ is the eigenfunction of eq.~(\ref{fluc}) 
(see eqs.~(\ref{eq:schpot}) and (\ref{eq:POT}), and note that $V''(\phi_{s}(x))=V_{s}(x)$), 
and $c_{m,n}$ are the expansion coefficients.  
Then the partition function becomes 
\begin{eqnarray*}
Z&\simeq&\exp\left(-S_{0}/\hbar\right) 
\int \prod_{m,n} {\rm d}c_{m,n} 
\int {\cal D}X(\tau) \int {\cal D}\Lambda(\tau)  \\
&\times&\exp\left[-\frac{1}{2}\frac{C}{\hbar}
\left(
\mid c_{m,n} \mid^2 \left(\omega_{m}^2 + \lambda_{n}\right) 
 - i 2 c_{m,n} \Lambda(\omega_{m}) \Phi_{n}(x_{i}) 
+ i 2 \Lambda(\omega_{m}) X(-\omega_{m}) 
\right)\right],
\end{eqnarray*}
where
\begin{eqnarray*}
\Lambda(\omega_{m})&=&\int {\rm d}\tau \Lambda(\tau) \exp(i \omega_{m} \tau), \\
X(\omega_{m})&=&\int {\rm d}\tau X(\tau) \exp(i \omega_{m} \tau).
\end{eqnarray*}
Twice Gaussian integrations bring   
\begin{eqnarray*}
Z&\simeq&\exp\left(-S_{0}/\hbar\right)  
\int {\cal D}X(\tau) \int {\cal D}\Lambda(\tau)  \nonumber \\
&\times&\exp\left[-\frac{1}{2}\frac{C}{\hbar}
\sum_{m,n}\left( 
\frac{\Lambda(\omega_{m})\Lambda(-\omega_{m})}{\omega_{m}^2 + \lambda_{n}}
\mid \Phi_{n}(x_{i}) \mid^2
+ i 2 \Lambda(\omega_{m}) X(-\omega_{m}) 
\right)\right] \nonumber \\
&\simeq&\exp\left(-S_{0}/\hbar\right)  
\int {\cal D}X(\tau)  
\exp\left[-\frac{1}{2}\frac{C}{\hbar}\sum_{m} 
\frac{X(\omega_{m})X(-\omega_{m})}
{\sum_{n}\frac{\mid \Phi_{n}(x_{i}) \mid^2}{\omega_{m}^2 + \lambda_{n}}}
\right].
\label{eq:hojopatfin}
\end{eqnarray*}
Here unimportant prefactors are neglected. 
Then the harmonic terms of the effective action is given by 
\begin{equation}
S_{\rm harm}=\frac{1}{2}C\sum_{m} 
\frac{X(\omega_{m})X(-\omega_{m})}
{\sum_{n}\frac{\mid \Phi_{n}(x_{i}) \mid^2}{\omega_{m}^2 + \lambda_{n}}}.
\label{eq:Hareff}
\end{equation} 

First we can now carry out the summation over $n$ in eq.~(\ref{eq:Hareff}). 
Equation~(\ref{eq:Hareff}) is expressed as a sum of 
the discrete eigenvalue part and the continuous 
one; i.e.
\begin{equation}
\sum_{n}\frac{\mid \Phi_{n}(x_{i}) \mid^2}{\omega_{m}^2 + \lambda_{n}}
=\frac{\mid \Phi_{0}(x_{i}) \mid^2}{\omega_{m}^2 + \lambda_{0}}
+\frac{L}{2 \pi} \int^{\infty}_{-\infty} {\rm d}k 
\frac{\mid \Phi(k;x_{i}) \mid^2}{\omega_{m}^2 + \lambda(k)}, 
\label{eq:sep}
\end{equation}
where $L$ is the system size. 
Note that the first-excited state solution, $\Phi_{1}(x)$, has odd parity
and does not appear in the above equation.  
In eq.~(\ref{eq:sep}), $k$ is a continuous 
parameter which characterizes the continuous eigenvalues instead of $n$. 
The integration of the second term can 
be carried out analytically with the approximation, 
$\Phi(k;x_{i}) \simeq \sqrt{1/L}\exp(ikx_{i})$. 
This is valid in the low temperature regime, which we are interested in, because 
the continuous states are higher energy states and give just a corrective 
contribution.   
Then 
\begin{eqnarray*}
\frac{L}{2 \pi} \int^{\infty}_{-\infty} {\rm d}k 
\frac{\mid \Phi(k;x_{i}) \mid^2}{\omega_{m}^2 + \lambda(k)}
&\simeq& \frac{1}{2 \pi} \int^{\infty}_{-\infty} {\rm d}k 
\frac{1}{\omega_{m}^2 + k^2 + \Delta_{\rm g}} \nonumber \\
&=&\frac{1}{2}\frac{1}{\sqrt{\omega_m^2+\Delta_{\rm g}}}. 
\label{eq:contint}
\end{eqnarray*}

Next we consider the summation over $\omega_m$. 
In order to obtain the tunneling rate, we calculate the bounce solution 
and substitute it 
into the action. The bounce solution has 
a characteristic time scale, $\omega_{\rm B}^{-1}$, 
which is shown in Fig.~\ref{fig:onebounce}. 
Then the main contribution to the summation over $\omega_m$ 
comes from terms with $\omega_{m} \simeq \omega_{\rm B}$. 
Now we expand $S_{\rm harm}$ into  
powers of $\omega_m$ up to the second order; 
\begin{equation}
\sum_{m} \frac{X(\omega_m)X(-\omega_m)}
{\frac{\mid \Phi_0(x_i) \mid^2}{\omega_m^2+\lambda_0}
+\frac{1}{2}\frac{1}{\sqrt{\omega_m^2+\Delta_{\rm g}}}}
\simeq \sum_{m}(M_0\omega_m^2 + K_0) X(\omega_m)X(-\omega_m), 
\label{eq:h}
\end{equation}
where
\[
M_0=\frac{\mid \Phi_0(x_i) \mid^2 + 
\frac{\lambda_0^2}{4\Delta_{\rm g}\sqrt{\Delta_{\rm g}}}}
{\left(\mid \Phi_0(x_i) \mid^2 + 
\frac{\lambda_0}{2\sqrt{\Delta_{\rm g}}}\right)^2}
\label{eq:mass0}
\]
and
\[
K_0=\frac{\lambda_0}{\mid \Phi_0(x_i) \mid^2 + 
\frac{\lambda_0}{2\sqrt{\Delta_{\rm g}}}}. 
\]
This approximation is relevant under 
the condition of $\omega_{\rm B}^2/\lambda_0 \ll 1$.  
Here $\omega_{\rm B}$ is the characteristic energy of the 
motion of the phase variable at the impurity site, $X(\tau)$, and 
$\lambda_0$ is that of the spatial fluctuation, $\Phi_0(x)$,  
around the ground state.  Then condition, $\omega_{\rm B}^2/\lambda_0 \ll 1$, 
means the spatial 
fluctuation around the ground state follows the motion of $X(\tau)$ 
adiabatically. 
This condition is checked 
self-consistently later. 

Finally we obtain the total effective action as 
\[  
S^{\rm eff}_{\rm total} = C\int {\rm d}\tau 
\left[ \frac{1}{2} 
M_0\left(\frac{\partial X}{\partial \tau}\right)^2 
+  U_{\rm imp}(X) \right],
\label{eq:aceff} 
\]
where $U_{\rm imp}(X)$ is the potential energy given by eq.~(\ref{eq:uimp}) 
with $X=\phi_i - \phi_ {s}(x_i)$. Here the 
harmonic term of the potential, $K_0 X^2$  in eq.~(\ref{eq:h}), 
is removed in order to avoid a double counting. 
We are interested in the tunneling 
near the threshold field, $\varepsilon_{\rm c}$,  
where we can expand $U_{\rm imp}$ into powers of 
$X$ and obtain 
\[
U_{\rm imp}(X) \simeq a_{\rm imp}X^2 - b_{\rm imp}X^3,  
\]
where 
\[
a_{\rm imp}=\frac{1}{2}
\left[-\frac{4 \left( -\varepsilon + \frac{1}{M}\sin(M \phi_{si})\right)}
{v\sin(\chi + \phi_{si})}+v\cos(\chi+\phi_{si})\right]  
\label{eq:ai}
\] 
and 
\[
b_{\rm imp}=\frac{1}{6}
\left[-\frac{\left(4 \left( -\varepsilon + \frac{1}{M}\sin(M \phi_{si})\right)\right)^2}
{\left(v\sin(\chi + \phi_{si})\right)^3}
+\frac{4\cos(M\phi_{si})}{v\sin(\chi + \phi_{si})}
+v\sin(\chi+\phi_{si})\right]. 
\label{eq:bi}
\] 
Here $\phi_{si} \equiv \phi_s(x_i)$, which is determined by eq.~(\ref{result}) 
for each field.  
Then the effective action near $\varepsilon_{\rm c}$ is 
\begin{equation}
S_{\rm eff} = C \int {\rm d}\tau 
\left[ \frac{1}{2} 
M_0
\left(\frac{\partial X}{\partial \tau}\right)^2 
+ a_{\rm imp}X^2 - b_{\rm imp}X^3 \right].
\label{eq:effthresh}
\end{equation}

\subsection{Quantum Tunneling Rate near the Threshold Field} 

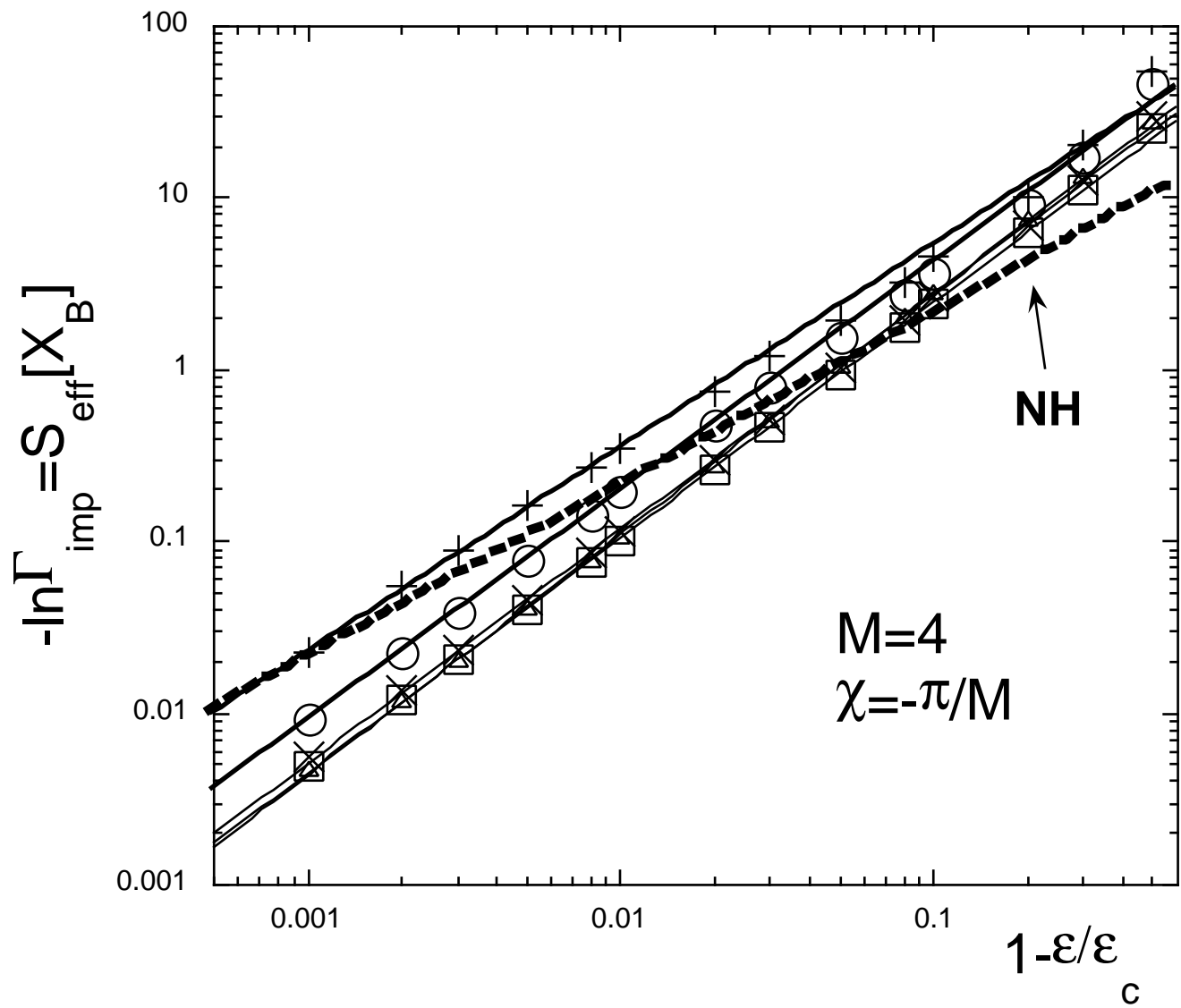
\begin{figure}[ht]
%\vspace{20cm}
%\epsfigure{file=figure6.eps,height=10cm}
\caption{The tunneling rates in the presence of the impurity  
for $M=4$ and $\chi=-\pi/M$. The electric field dependencies are different 
each other for various $v$ values. The dotted line shows the result of Nakaya and Hida. }
\label{fig:qtun}
\end{figure}

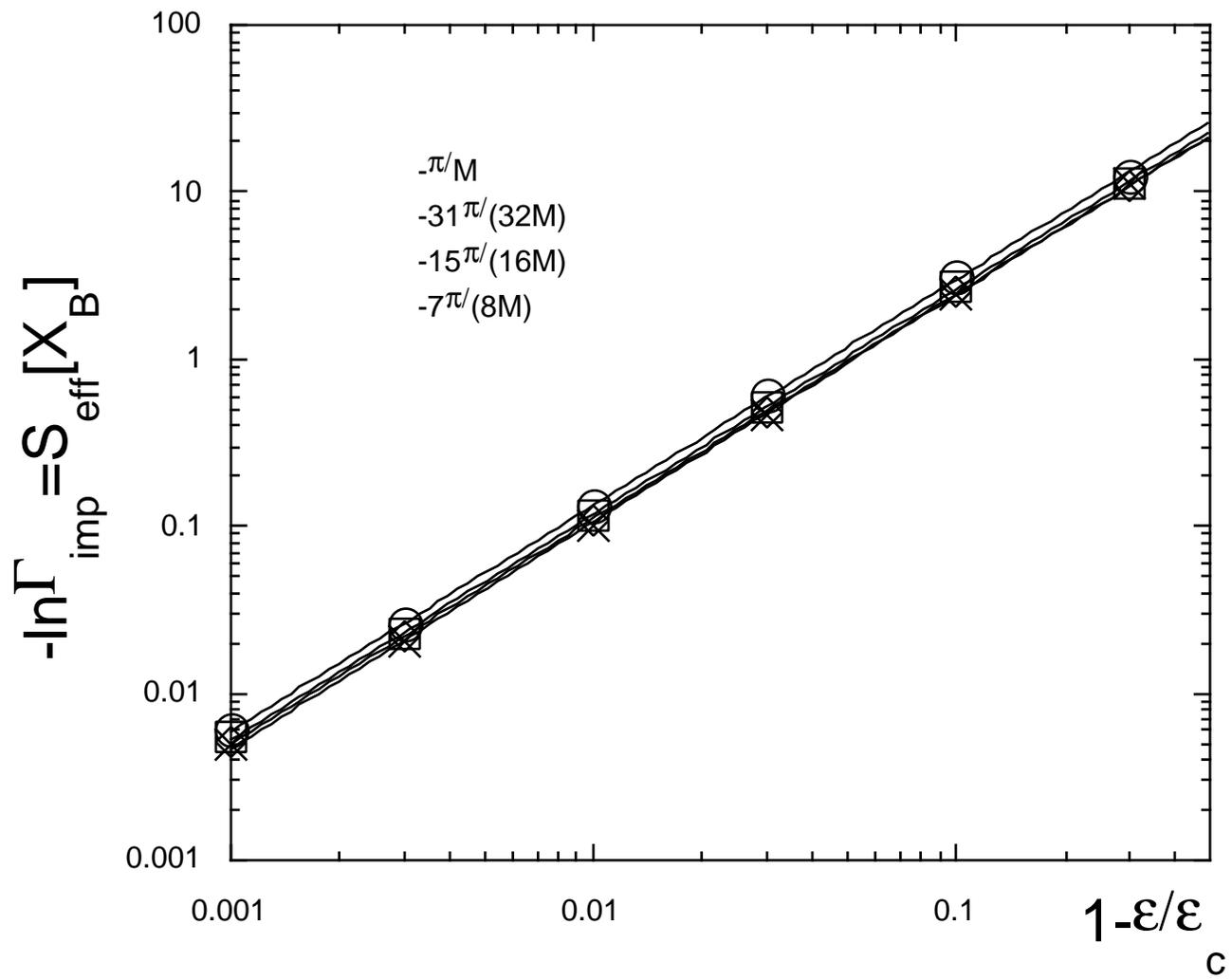
\begin{figure}[ht]
%\epsfigure{file=figure7.eps,height=10cm}
%\vspace{20cm}
\caption{The tunneling rates in the presence of the impurity for
 $M=4$ and $v=1$. The electric field dependencies are not so different each other 
for various $\chi$ values.}
\label{fig:qtunc}
\end{figure}

We calculate the tunneling rate 
using the effective action near the threshold field.  
The tunneling rate within the WKB approximation is given by
\[
\Gamma_{\rm imp} \propto 
\exp\left[-S_{\rm eff}[X_{\rm B}]/\hbar\right]
\]
where $X_{\rm B}(\tau)$ is the bounce solution of the equation of motion 
which is derived from $S_{\rm eff}$ as  
\[
-M_0\frac{\partial^2 X}{\partial \tau^2}+2a_{\rm imp}X-3b_{\rm imp}X^2
=0. 
\label{eq:motion}
\]
This equation of motion can be solved analytically and gives the solution as 
\[
X_{\rm B}(\tau)=\frac{A_{\rm B}}{\cosh^2(\omega_{\rm B}\tau)},
\label{eq:b} 
\]
where 
\begin{eqnarray*}
A_{\rm B}&=&\frac{a_{\rm imp}}{b_{\rm imp}}, \\
\omega_{\rm B}&=&  
\sqrt{\frac{a_{\rm imp}}{2 M_0}}. 
\end{eqnarray*} 
Substituting $X_{\rm B}(\tau)$ into $S_{\rm eff}[X]$ and 
carrying out the $\tau$-integration,  
we obtain the tunneling rate as  
\begin{equation}
\Gamma_{\rm imp} \propto 
\exp\left[ -\frac{C}{\hbar}
\frac{8}{15}\sqrt{2 M_0}
\frac{a^{\frac{5}{2}}_{\rm imp}}{b^2_{\rm imp}}\right]. 
\label{eq:tunimp}
\end{equation}
The obtained tunneling rates are shown in 
Figs.~\ref{fig:qtun} and~\ref{fig:qtunc}. Here we take the 
ratio of the effective mass of the CDW to the mass of the band electron 
as $m^{\ast}/m_{\rm b}=1 \times 10^3$. This parameter determines a value 
of $C$ (see eq.~(\ref{eq:ch})).   
The $v$-dependence of the tunneling rate is shown in 
Fig.~\ref{fig:qtun} for $M=4$ and $\chi=-\pi/M$,  
which gives the minimum value of the classical threshold 
field.  
On the other hand, the $\chi$-dependence of the tunneling rate is shown in 
Fig.~\ref{fig:qtunc} for $M=4$ and $v=1$. 
The condition, $\omega_{\rm B}^2/\lambda_0 \ll 1$, that means the present 
effective action is relevant, 
is satisfied 
in all the region in Figs.~\ref{fig:qtun} and~\ref{fig:qtunc}.   
We also note that the WKB approximation is relevant in the region of 
$-\ln\Gamma_{\rm imp}=S_{\rm eff}[X_{\rm B}]/\hbar  < 1$.   

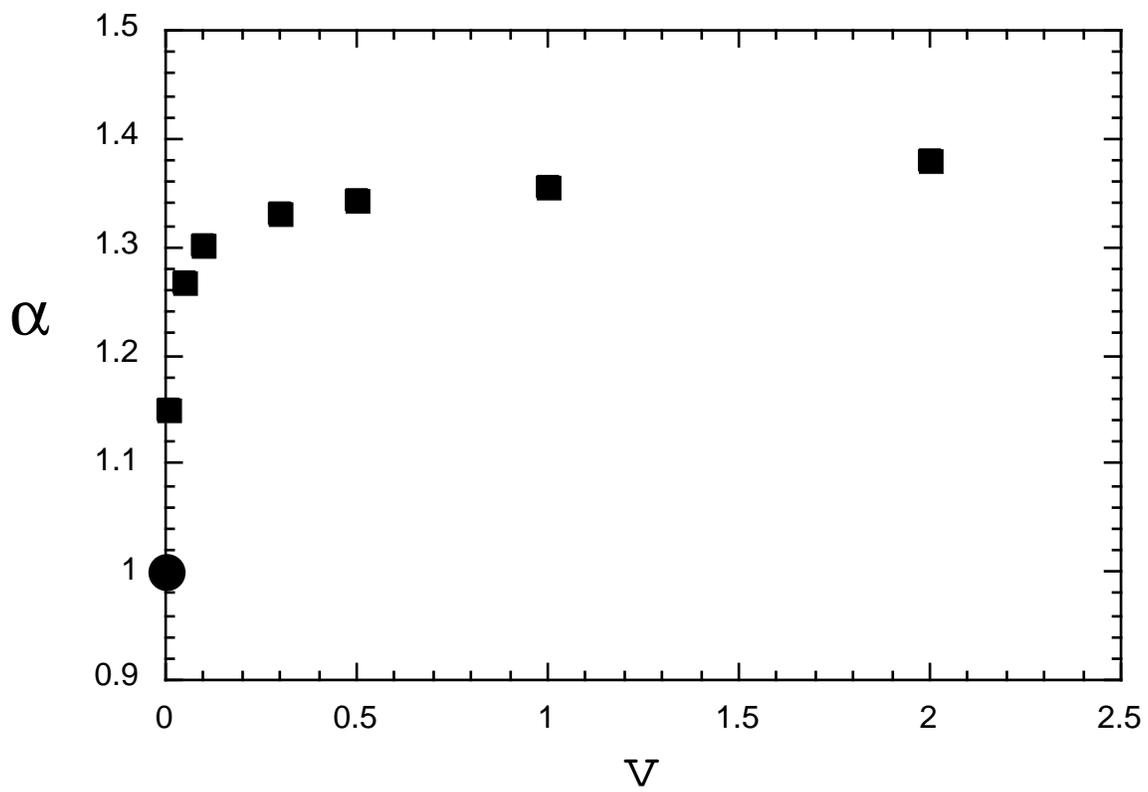
\begin{figure}[ht]
%\vspace{20cm}
%\epsfigure{file=figure8.eps,height=10cm}
\caption{The $v$-dependence of $\alpha$ for $M=4$ and $\chi=-\pi/M$. 
As $v$ is getting smaller,  
$\alpha$ becomes smaller.  When $v=0$, $\alpha(0)=1$, which is
given by Nakaya and Hida by the scaling argument. When $v$ exceeds $2$ 
in the case of $\chi=-\pi/M$ ($M=4$), the present calculation 
gradually becomes unreliable. } 
\label{fig:expt}
\end{figure}

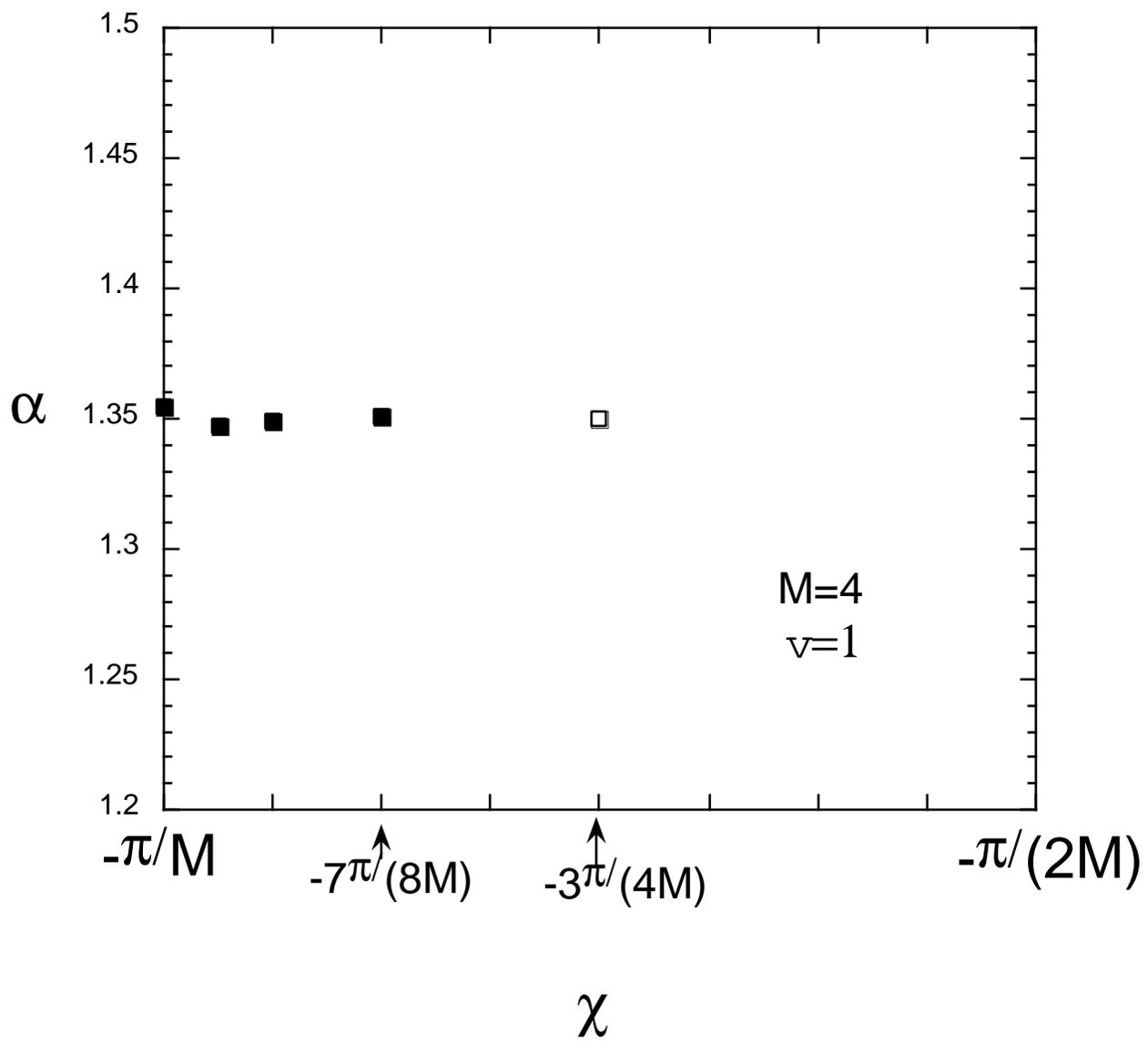
\begin{figure}[ht]
%\epsfigure{file=figure9.eps,height=10cm}
%\vspace{20cm}
\caption{The $\chi$-dependence 
of $\alpha$ for $M=4$ and $v=1$. 
In this case,  $\alpha$ does not depends on $\chi$ 
strongly. When 
$\chi$ exceeds $-7\pi/(8M)$ ($M=4$) in the case of $v=1$, the present calculation  
gradually becomes unreliable. Note that the left most data point (open square) 
should be considered 
as semiquantitative one because of irrelevancy of the approximation we took (see 
text). } 
\label{fig:expt2}
\end{figure}

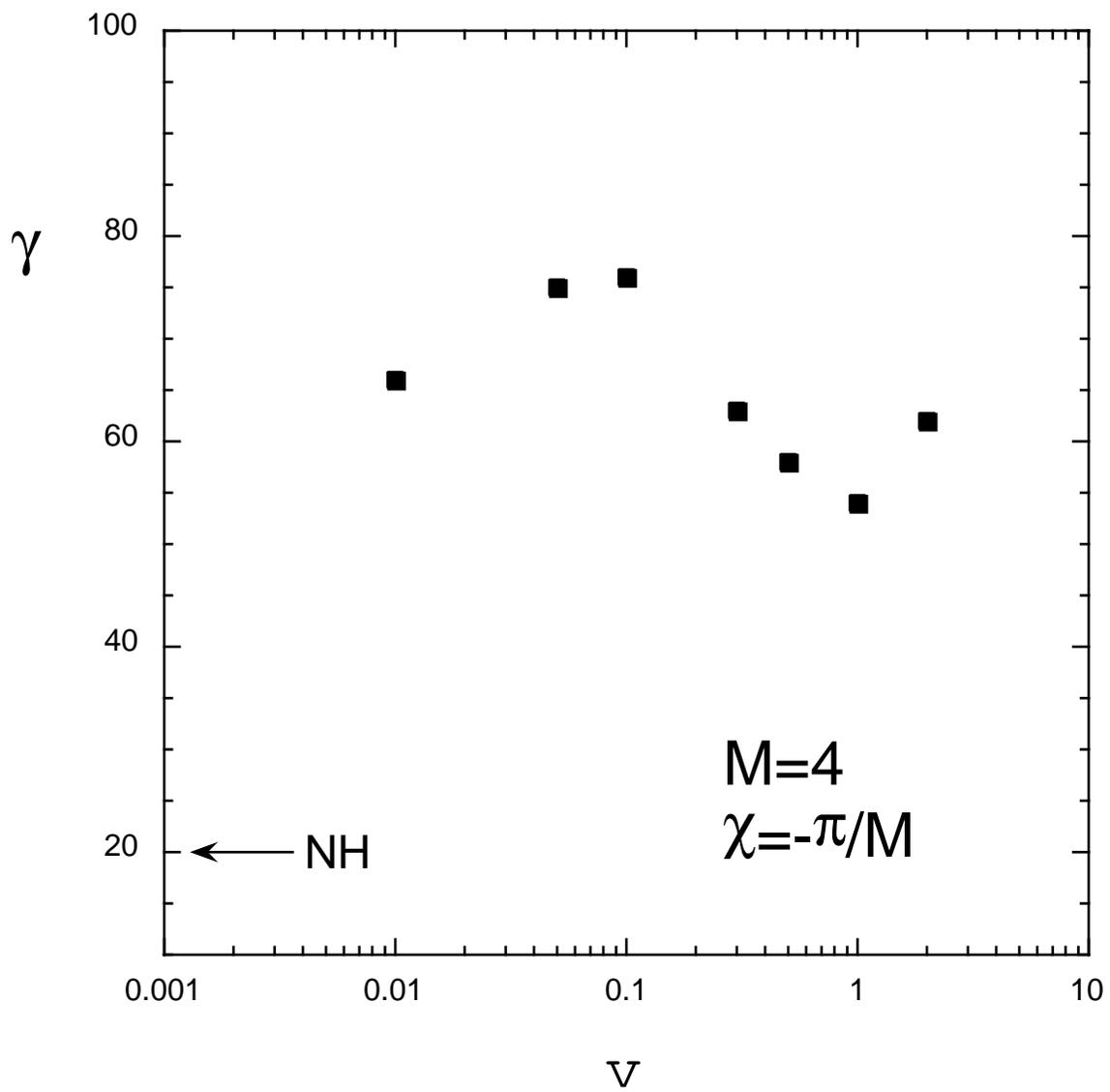
\begin{figure}[ht]
%\epsfigure{file=figure10.eps,height=10cm}
%\vspace{20cm}
\caption{The $v$-dependence 
of $\gamma$ for $M=4$ and $\chi=-\pi/M$. 
Values of $\gamma$ distribute  
around the value, $ 6\times 10^1$. }
\label{fig:gammav}
\end{figure}

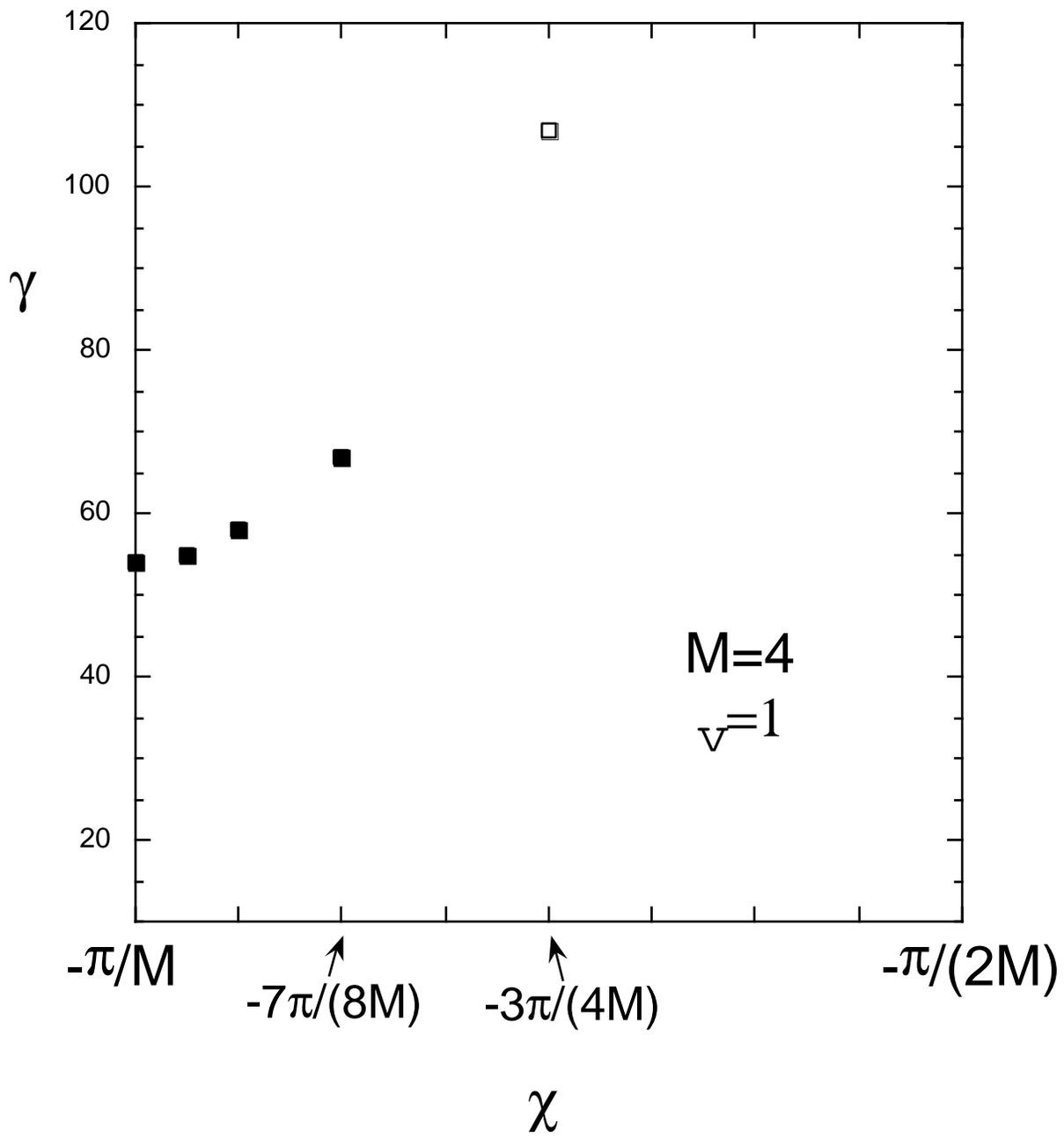
\begin{figure}[ht]
%\epsfigure{file=figure11.eps,height=10cm}
%\vspace{20cm}
\caption{The $\chi$-dependence 
of $\gamma$ for $M=4$ and $v=1$.  
The value of $\gamma$ 
monotonically increases as $\chi$ decreases.  
Note that the left most data point (open square) 
should be considered 
as semiquantitative one because of irrelevancy of the approximation we took (see 
text). } 
\label{fig:gammachi}
\end{figure}

The results shown in Figs.~\ref{fig:qtun} and~\ref{fig:qtunc} 
can be expressed as 
\begin{equation}
\Gamma_{\rm imp} \propto \exp\left[-\gamma (v,\chi ) 
\left(1-\frac{\varepsilon}{\varepsilon_{\rm c}(v,\chi)}\right)^{\alpha(v,\chi)}\right]. 
\label{eq:expa}
\end{equation} 
Here $\alpha$ is a kind of generalized critical exponent and 
$\gamma$ is some factor. The $v$ and $\chi$-dependencies  
of $\alpha$ are  
shown in Figs.~\ref{fig:expt} and~\ref{fig:expt2}, respectively. 
When $v$ exceeds $2$ 
in the case of $\chi=-\pi/M$ ($M=4$) and 
$\chi$ exceeds $-7\pi/(8M)$ ($M=4$) in the case of $v=1$, either of 
the two conditions corresponding to the relevances of the present effective action, 
$\omega_{\rm B}^2/\lambda_0 \ll 1$, and of the WKB approximation  
gradually becomes invalid at any electric field. 
This means the present effective model 
is relevant in a limited region in the parameter space.  
Dependencies of $\gamma$ on $v$ and $\chi$ are shown in Fig.~\ref{fig:gammav} 
and Fig.~\ref{fig:gammachi}.    
In the relevant region, $\alpha$ depends on $v$ strongly, but does not on $\chi$. 

In the absence of the impurity, an effective theory near the threshold field 
was  given by Nakaya and Hida (NH) based on the scaling argument.~\cite{rf:Nakaya}
The effective action of this theory is obtained by expanding $U(\phi)$, eq.~(\ref{eq:U}),  
at the inflection point, which yields  
This effective action is scaled as 
\[
S_{\rm E-eff}=C\left(\frac{a_{1}}{b_{1}}\right)^2 
\int {\rm d}\tilde{\tau} {\rm d}\tilde{x} 
\left[\frac{1}{2}\left(\frac{\partial \tilde{\phi}}{\partial \tilde{\tau}}
\right)^2
+\frac{1}{2}\left(\frac{\partial \tilde{\phi}}{\partial \tilde{x}}\right)^2
+\tilde{\phi}^2 -\tilde{\phi}^3\right]
\label{eq:sNHaction}
\]
where $\tilde{\phi}=(a_{1}/b_{1})\phi_{\rm eff}$, 
$\tilde{\tau}=\sqrt{a_{1}}\tau$, $\tilde{x}=\sqrt{a_{1}}x$,  
$\phi_{\rm eff}=\phi - \pi/(2M) 
+ (2/M)\sqrt{(1-\varepsilon/\varepsilon_{\rm T})/2}$,@
$a_{1} \equiv \sqrt{(1-\varepsilon/\varepsilon_{\rm T})/2}$, 
and $b_{1} \equiv (1/3!) M$. 
. 
Therefore we obtain the electric field dependence of the tunneling rate, $\Gamma_0$, 
without any knowledge about the bounce solution as 
\[
\Gamma_0 \propto \exp\left[-B_0 
\left(1-\frac{\varepsilon}{\varepsilon_{\rm T}}\right)^{1}\right]. 
\label{eq:NHscale}
\]
NH evaluated $B_0$ numerically as 
\[
B_0 \simeq \frac{C}{\hbar} \frac{6.2\times 10}{M^2},  
\label{eq:NHresult}
\]
in the present notation. 
We compare the present result with NH's. 
In the limit of vanishing impurity potential, $\alpha$ in eq.~(\ref{eq:expa}) 
tends to unity 
which was given by NH in the absence of the impurity. 
On the other hand $\gamma$ in the present model, around $6 \times 10^1$, 
is about three times larger than that given by NH 
($B_0 \simeq 2.0 \times 10^1$ 
for the present parameter.).

We conclude that in the presence of the 
impurity $\alpha$ is larger than unity
in the absence of the impurity.  
In the small limit of $v$, however, $\alpha$ becomes close to unity. 
Values of $\gamma$ given by the present calculation is about three times larger than 
that given by NH. Then present results in the 
small $v$ limit shows the consistency with those of NH qualitatively, however, 
does not show exact coincidence quantitatively.  
The present effective model is applicable when the localized fluctuation can follow the motion of the phase 
at the impurity site, $X(\tau)$, adiabatically. 
Then the origin of the $v$-dependence of $\alpha$ seems to
come from the existence of the localized fluctuation over the length scale 
of $\xi$ around the impurity,  
as in the case of the lowering of the classical threshold field.~\cite{rf:Yumoto} 
We also conclude that the impurity can enhance the tunneling rate near the 
threshold field, because 
$\alpha >1$ in that case. 
According to the scaling argument by NH,~\cite{rf:Nakaya} the lower dimensional 
system has larger $\alpha$:  spatial dimension = 1, 2 and 3 have 
$\alpha$ =1, 3/4 and 1/2, respectively. Then $\alpha >1$ means that 
the impurity brings the 1D system 
to the {\it lower-dimensional} system effectively.

\section{Conclusion and Discussions}

We have investigated the quantum depinning of the commensurate CDW 
with one impurity at absolute zero 
temperature theoretically.  
The impurity causes the different 
electric field dependence of the tunneling rate from 
that in its absence and enhances the quantum depinning. 
In this issue, the fluctuation around the ground state plays 
an important role. It is described by the eigenvalue equation~(\ref{fluc}). 
In the absence of the impurity the eigenvalue equation has no bound state. On the other 
hand, in the presence of the impurity under 
the condition of $-\pi/M \leq \chi < -\pi/(2M)$
the eigenvalue equation has a bound state, which
means there is a spatially localized fluctuation around the impurity. 
This localized fluctuation which is induced by the impurity 
is the origin of the local sliding around the 
impurity and assists the depinning.  

Finally we discuss future problems. 
The model which has been treated in the present paper offers some 
fundamental understandings of the effects of 
impurities on the depinning of the commensurate CDW. 
However, it is an idealized model. 
In actual cases, the following three points are important; 
dissipation, a finite density of impurities and three dimensionality. 
The dissipation is neglected in the present model. 
In the model once the local sliding occurs, the sliding expands to the whole 
system by the energy gain from the electric field. 
If a strong dissipation exists, however, the local sliding will be affected. 
Realistic systems contain not only 
one impurity but a finite density of impurities. 
However, the present result is expected to be applicable to the commensurate CDW with dilute but 
with macroscopic numbers of impurities where 
the inverse of the impurity density is larger than the 
phase coherence length, $\xi$.   
The depinning will be triggered by 
the local sliding at the optimal impurity site, 
$\chi = -\pi/M$.  
The effects of the high density of impurities is remained as a future problem. 
In one-dimensional systems, the 
impurity, a zero-dimensional object, has stronger effects 
than in higher dimensional systems. 
Generalization of the present model to three dimensional systems 
is left for future studies. 

\section*{Acknowledgments}
We would like to thank Hiroshi Kohno for fruitful discussions. 
The present work is a part of Doctor Thesis of one of the authors (M.Y). 
He is grateful for precious advice given by members of the committee for 
judgment: Hiroshi Fukuyama, 
Masao Ogata, Hajime Takayama, Seigo Tarucha and Miki Wadati 
(University of Tokyo).

\newpage
\section*{List of Figures}

\begin{figure}[ht]
%\vspace{10cm}
%\epsfigure{file=figure5.eps,height=10cm}
\caption{One bounce solution. Its typical time scale 
is $\omega_{\rm B}^{-1}$. }
\label{fig:onebounce}
\end{figure}

\begin{thebibliography}{99}
\bibitem{ref:note1}
The threshold field, $\varepsilon_{\rm c}$, is given by 
the solution of the differential equation (\ref{fluc}) for $\lambda=0$, 
but in this case  
the differential equation can be 
transformed to algebraic equations by the method shown in 
ref. 4.
\end{thebibliography}
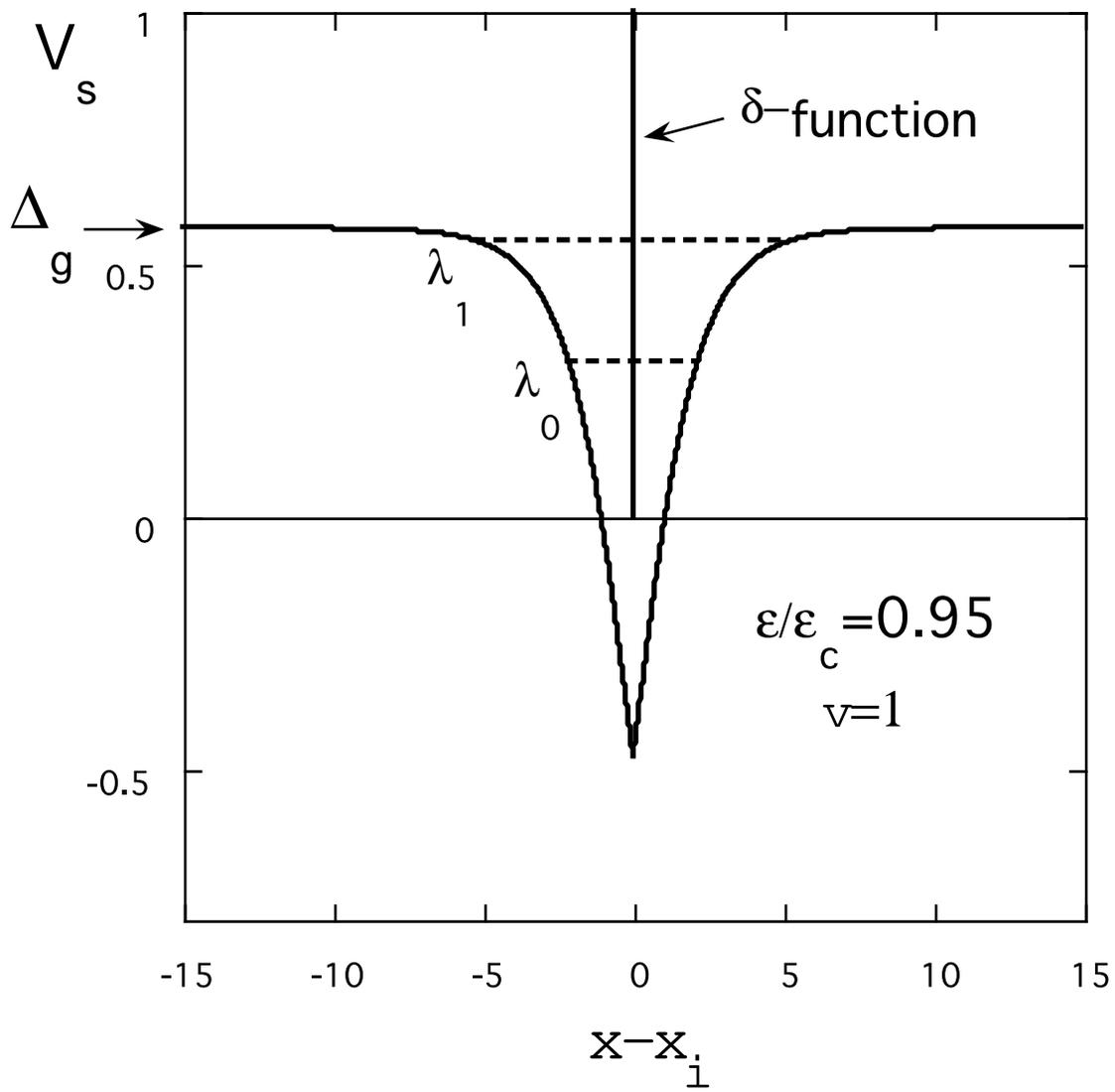
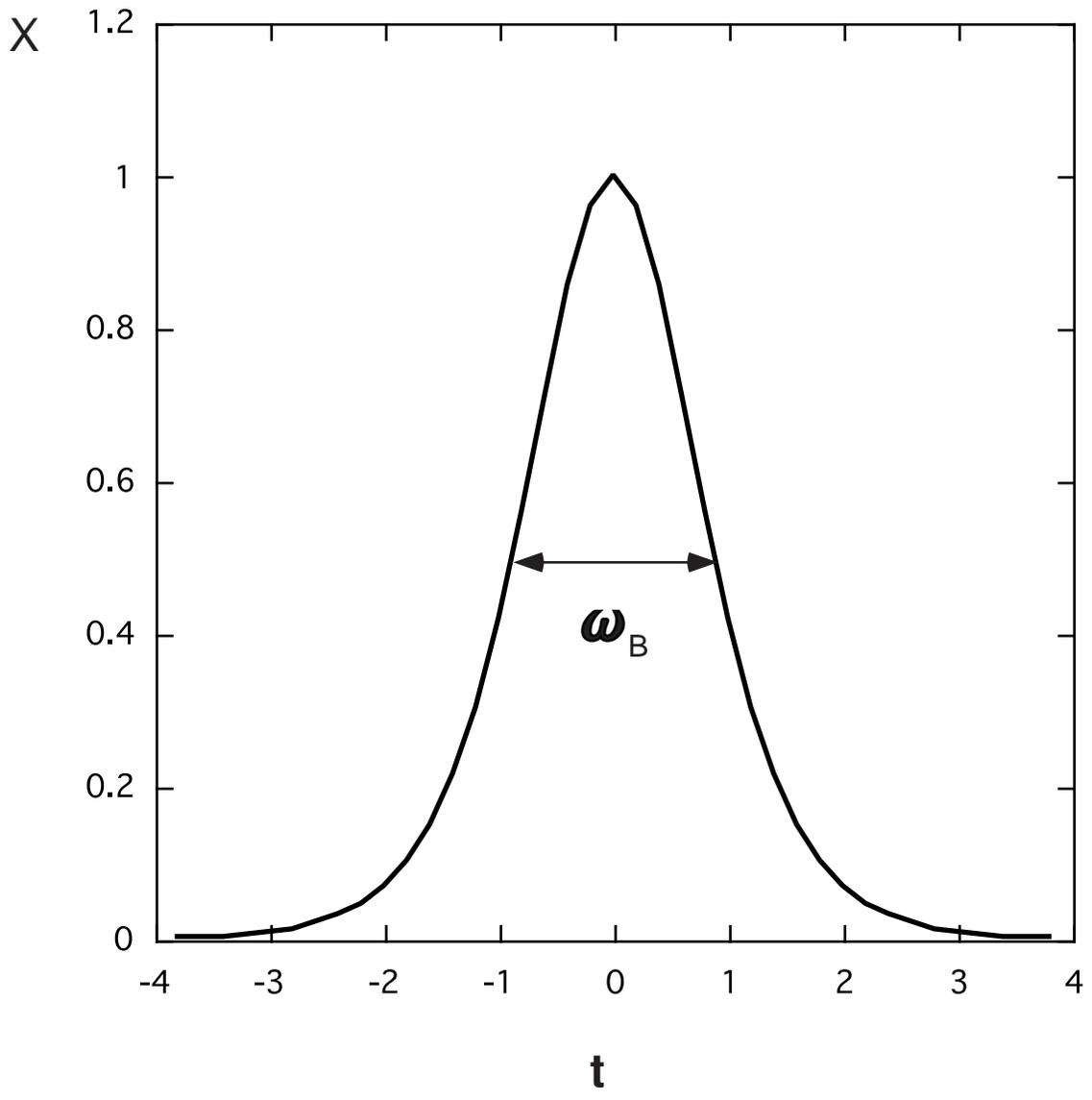
\end{document}